\documentstyle[preprint,aps,prabib]{revtex}
\begin{document}
\draft
\title{A Single Classical Quark}
\author{V. Dzhunushaliev
\thanks{E-mail address: dzhun@freenet.bishkek.su}}
\address{Theoretical Physics Department,\\
the Kyrgyz State National University,\\ 
720024, Bishkek, Kyrgyzstan}
\date{}
\maketitle
\begin{abstract}
The spherically symmetric solution in classical $SU(3)$ 
Yang - Mills theory is found. It is supposed that such 
solution describes a classical quark. It is regular in 
origin and hence the interaction between two quarks 
is small on the small distance. The obtained solution has the 
singularity on infinity. It is possible that is the reason 
why the free quark cannot exist. Evidently, nonlocality of 
this object leads to the fact that in quantum chromodynamic 
the difficulties arise connected with investigation of quarks 
interaction on large distance.
\end{abstract}
\narrowtext

\section{Introduction}

The some properties of the classical electron have the 
significant meaning in quantum electrodynamic. For 
example, an electron is the point source of electrical 
field, and this make possible to interpret a propagator 
as propagation function of point electron from initial 
point to endpoint. Thus, the classical properties of 
elementary particles have the important sense in quantum 
field theory.
\par 
In this work we obtain the spherically symmetric solution 
of the classical $SU(3)$ Yang - MIlls equations. This solution 
we name 
as a ``quark'' (the quotes signs that some degrees of freedom 
are omit). This solution is regular in origin and this can means 
that in quantum chromodynamic the quarks weakly interact 
on small distance. On the other hand this solution has 
singularity on infinity, it is possible that this property 
means that a solitary quark cannot exists.
\par
The spherically symmetric solutions for $SU(2)$ classical 
Yang - Mills theory had been investigated in Ref's \cite{1}. 
In these papers was shown that these solutions have singularities 
in origin and in some distance from origin.

\section{The initial equations}

The ansatz for the $SU(3)$ gauge field we take as in \cite{2}:
\begin{mathletters}
\label{1}
\begin{eqnarray}
A^a & = & \frac{2\varphi(r)}{Ir^2} \left( \lambda ^2 x - \lambda ^5 y 
      + \lambda ^7 z\right ) + \frac{1}{2}\lambda ^a
      \left( \lambda ^a _{ij} + \lambda ^a_{ji} \right ) 
      \frac{x^ix^j}{r^2} w(r),
\label{1a:1}\\
A^a_i & = & \left( \lambda ^a_{ij} - \lambda ^a_{ji} \right )
        \frac {x^j}{Ir^2} \left(f(r) - 1\right ) +
        \lambda ^a_{jk} \left (\epsilon _{ilj} x^k + 
	\epsilon _{ilk} x^j\right ) \frac{x^l}{r^3} v(r)
\label{1b:2},
\end{eqnarray}
\end{mathletters}
here $\lambda ^a$ are the Gell - Mann matrixes; $a=1,2,\ldots ,8$
is color index; Latin indexes $i,j,k,l=1,2,3$ are the space indexes; 
$I^2=-1$; $r, \theta, \varphi$ are the spherically coordinate system. 
Substituting Eq's (\ref{1}) in the Yang - Mills equations:
\begin{equation}
\frac{1}{\sqrt{-g}}\partial _{\mu} \left (\sqrt {-g} {F^{a\mu}}_{\nu}
\right ) + f^{abc} {F^{b\mu}}_{\nu} A^c_{\mu} = 0,
\label{2}
\end{equation}
we receive the following $SU(3)$ equations system for 
$f(r), v(r), w(r)$ and $\varphi (r)$ functions:
\begin{mathletters}
\label{3}
\begin{eqnarray}
r^2f''& =& f^3 - f + 7fv^2 + 2vw\varphi - f\left (w^2 + \varphi ^2\right ),
\label{3a:1}\\
r^2v''& = & v^3 - v + 7vf^2 + 2fw\varphi - v\left (w^2 + \varphi ^2\right ),
\label{3b:2}\\
r^2w''& = & 6w\left (f^2 + v^2\right ) - 12fv\varphi,
\label{3c:3}\\
r^2\varphi''& = & 2\varphi\left (f^2 + v^2\right ) - 4fvw.
\label{3d:4}
\end{eqnarray}
\end{mathletters}
This set of equations is very difficult even for numerical
investigations. We will investigate a more simpler case when only
two functions are nonzero. It is easy to see that there can be only three 
cases. The first case is well-known monopole case by $(f,w=0)$
or $(v,w=0)$. Here we will investigate only $v,w \neq 0$ case.

\section{The quark solution}

Here we examine $f=\varphi=0$ case. The case $v=\varphi=0$ is
analogous. Now the input equations have the following form:
\begin{mathletters}
\label{9}
\begin{eqnarray}
r^2 v''& = & v^3 - v - vw^2,
\label{9:1}\\
r^2 w''& = & 6w v^2.
\label{9:2}
\end{eqnarray}
\end{mathletters}
We seek the regular solution near $r=0$ point. The Eq's (\ref{9}) 
demand that $v$ and $w$ functions have the following view at origin
$r=0$:
\begin{mathletters}
\label{10}
\begin{eqnarray}
v & = & 1 + v_2 \frac{r^2}{2!} + \ldots,
\label{10:1}\\
w & = & w_3\frac{r^3}{3!} + \ldots .
\label{10:2}
\end{eqnarray}
\end{mathletters}
The numerical integration of Eq's (\ref{9}) is displayed on Fig.1,2. 
The asymptotical behaviour of received solution $(r\to \infty )$
is as follows:
\begin{mathletters}
\label{11}
\begin{eqnarray}
v & \approx & a \sin \left (x^{1+\alpha } + \phi _0\right ),
\label{11:1}\\
w & \approx & \pm\left [ (1 + \alpha ) x^{1 + \alpha } + 
\frac{\alpha}{4}\frac{\cos {\left (2x^{1 + \alpha} + 2\phi _0 \right )}}
{x^{1 + \alpha}}\right ],
\label{11:2}\\
3a^2 & = & \alpha(\alpha + 1).
\label{11:3}
\end{eqnarray}
\end{mathletters}
here $x=r/r_0$ is dimensionless radius; $r_0, \phi _0$ are some 
constants. For our potential $A^a_{\mu}$ we have the following 
nonzero color ``magnetic'' and ``electric'' fields: 
\begin{mathletters}
\label{12}
\begin{eqnarray}
H^a_{\varphi} & \propto & v',
\label{12:1}\\
H^a_{\theta} & \propto & v',
\label{12:2}\\
E^a_r & \propto & \frac{rw' - w}{r^2},
\label{12:3}\\
E^a_{\varphi} & \propto & \frac{vw}{r},
\label{12:4}\\
E^a_{\theta} & \propto & \frac{vw}{r},
\label{12:5}\\
H^a_r & \propto & \frac{v^2 - 1}{r^2},
\label{12:6}
\end{eqnarray}
\end{mathletters}
here for Eq's (\ref{12:1}), (\ref{12:2}) and (\ref{12:3}) the color index 
$a=1,3,4,6,8$ and for Eq's (\ref{12:4}), (\ref{12:5}) and (\ref{12:6}) 
$a=2,5,7$. Analyzing the asymptotical behaviour of the 
$H^a_{\varphi}, H^a_{\theta}, H^a_r$ and $E^a_{\varphi}, E^a_{\theta}$ 
fields we see that they are the strongly oscillating fields.
It is interesting that the radial components of the ``magnetic''
and ``electric'' fields drop to zero variously at infinity:
\begin{mathletters}
\label{13}
\begin{eqnarray}
H^a_r & \approx & \frac{1}{r^2},
\label{13:1}\\
E^a_r & \approx & \frac{1}{r^{1-\alpha}}.
\label{13:2}
\end{eqnarray}
\end{mathletters}
Among all the (\ref{12}) fields only the radial components of 
``electric'' fields are nonoscillating. From Eq's (\ref{11}) we see
that our solution has the oscillating part (\ref{11:1}) and confining
potential (\ref{11:2}). It is necessary to mark that obtained solution
is solution with arbitrary initial condition, whereas the monopole
solution is a special case of the initial condition. 
The expression for an energy density has the following view: 
\begin{equation}
\epsilon \propto 4\frac{v'^2}{r^2} + \frac{2}{3}\left (
\frac{w'}{r} - \frac{w}{r^2}\right )^2 + 4\frac{v^2w^2}{r^4} +
\frac{2}{r^2} \left (v^2 - 1\right )^2.
\label{14}
\end{equation}
This function is displayed on the Fig.3.
\par
We note the asymptotical form of the energy density followed
from (\ref{12}) condition:
\begin{equation}
\epsilon \approx \frac{2}{3} \frac{\alpha (1 + \alpha)^2 (3\alpha + 2)}
{x^{2-2\alpha }}.
\label{15}
\end{equation}
In the first approximation $\epsilon$ is nonoscillating on the infinity.
The asymptotical form of $\epsilon$ leads to that the energy of
such solution is infinity. 

\section{Discussion}

What is the physical meaning of this solution? It is possible that 
it is analogous to the Coulomb potential in electrostatics. 
But an electron can exist in empty space while a quark is not
observable in a free state. Therefore, the obtained solution can  
describe the \underline{classical} color charge - ``quark''. 
The quotation marks indicate that we examine the simplified 
Eq's (\ref{9}) instead of complete Eq's (\ref{3}).
We see that classical electron and ``quark'' 
have the fundamental difference among 
themselves. The electron has a singularity at origin by $r=0$, 
but  the ``quark'' at infinity by $r=\infty$. It should be also 
noted that this solution has the asymptotical freedom property 
since at origin $r=0$ the gauge potential $A^a_{\mu} \to const$.
\par
Thus, we suppose that the received here solution have the physical
significance: the static spherically symmetric solution 
of $SU(3)$ Yang - Mills equations describes a \underline{classical} 
single ``quark''.

\newpage
\centerline{List of figure captions}
\par
Fig.1. Oscillating potential $v(r)$
Fig.2. Potential $w(r)$.
Fig.3. Energy density $\epsilon (r)$.

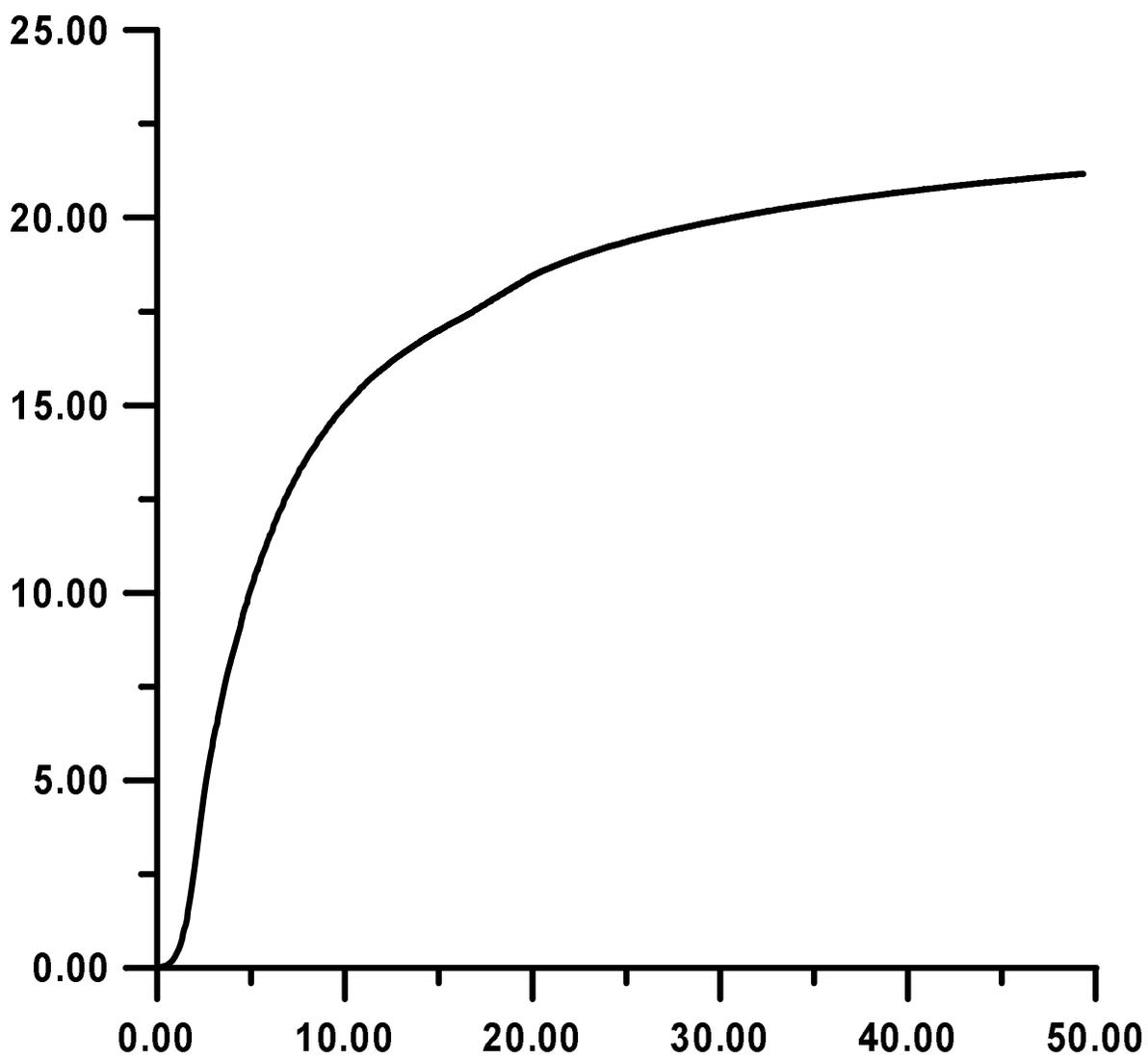
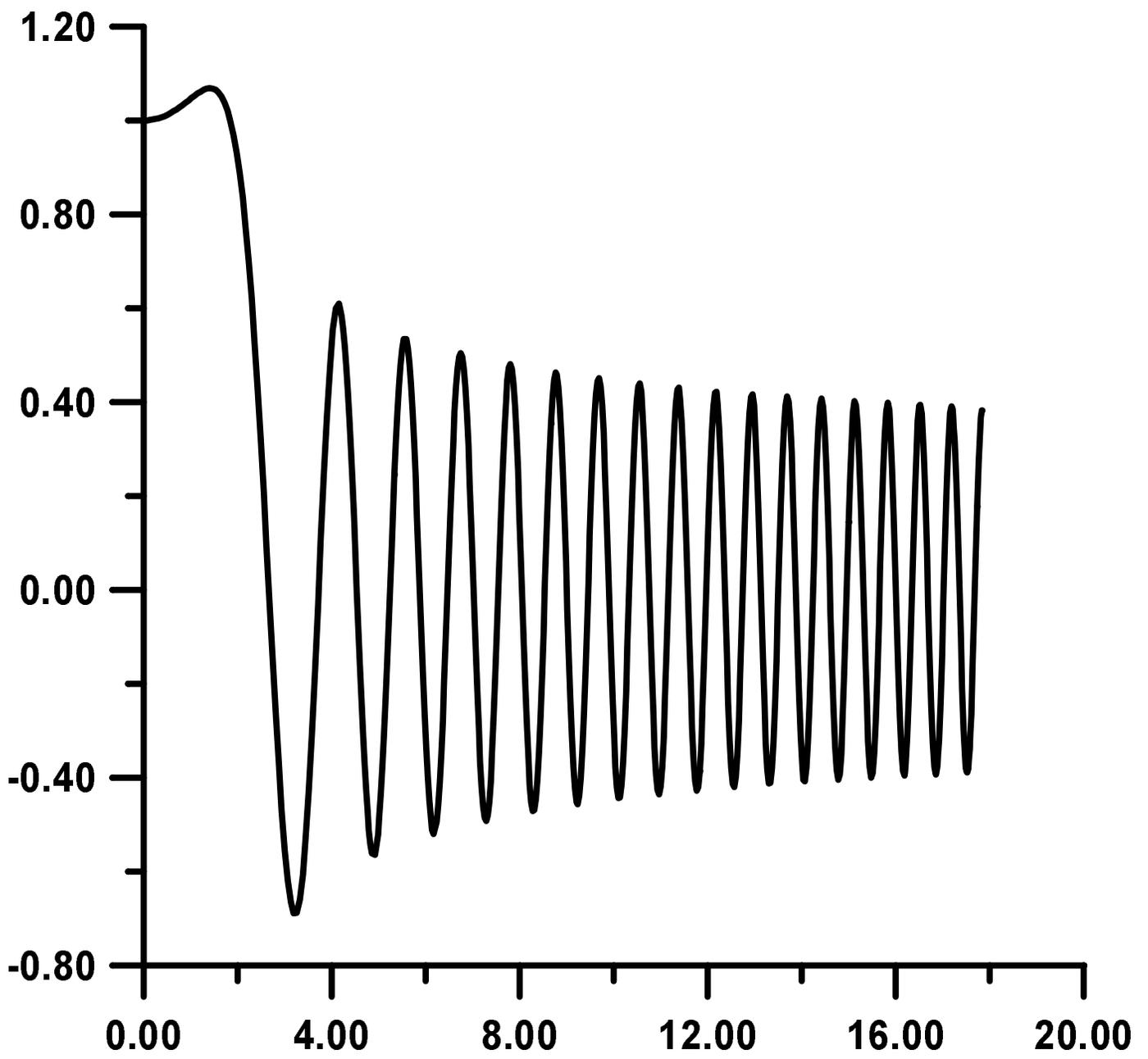
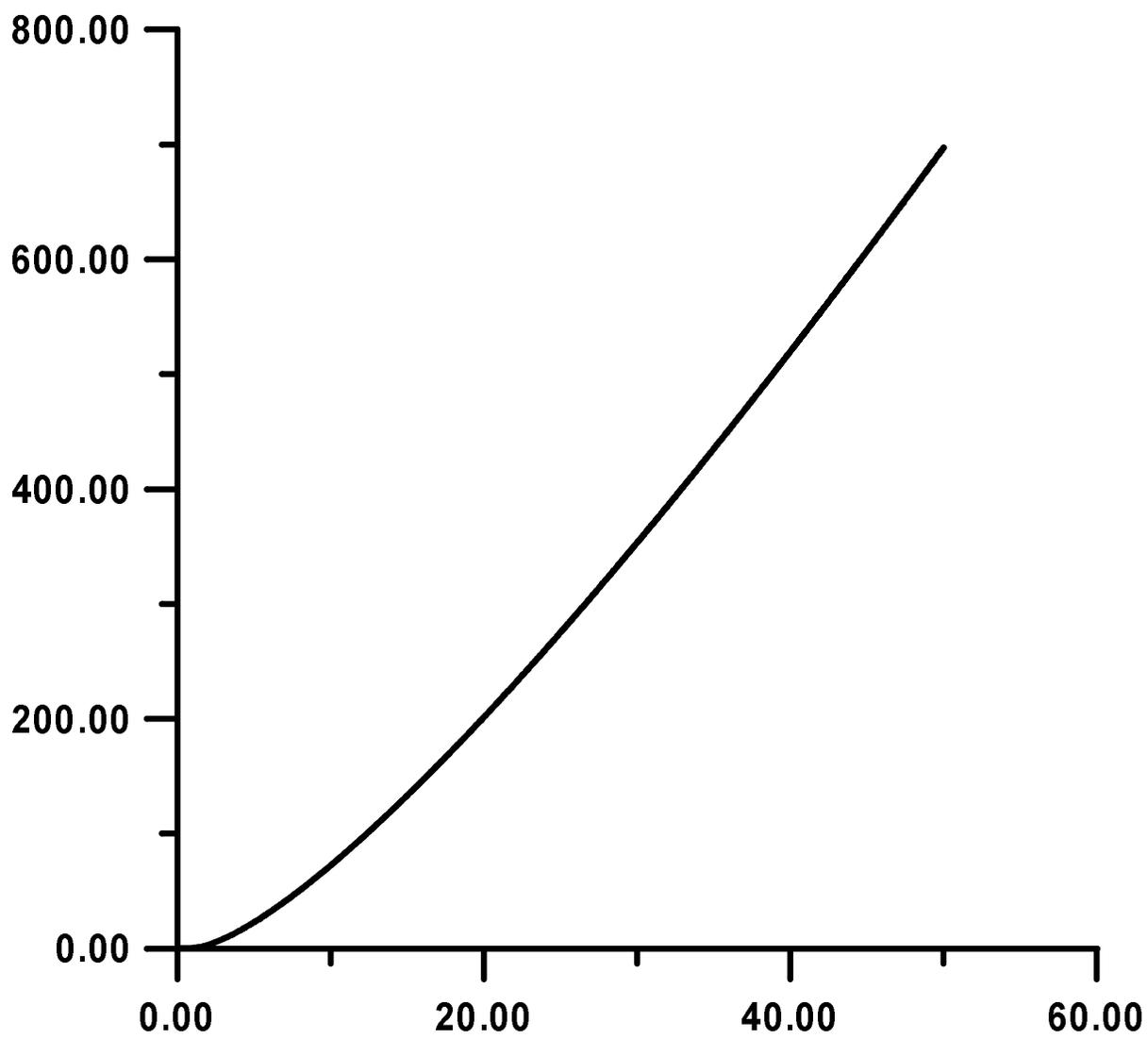
\end{document}